\documentclass[twoside,slac_one]{revtex4}
\usepackage{graphicx}
\usepackage{fancyhdr}
\usepackage{amsmath} 
\usepackage{bm}
\usepackage{amsxtra}
\usepackage{amssymb}
\usepackage{amsthm}
\usepackage{latexsym}
\usepackage{lscape}

\begin{document}

\newcommand{\dzero}     {D\O}
\newcommand{\ppbar}     {\mbox{$p\bar{p}$}}
\newcommand{\ttbar}     {\mbox{$t\bar{t}$}}
\newcommand{\bbbar}     {\mbox{$abb\bar{b}$}}
\newcommand{\ccbar}     {\mbox{$c\bar{c}$}}
\newcommand{\pythia}    {{\sc{pythia}}\xspace}
\newcommand{\alpgen}    {{\sc{alpgen}}\xspace}
\newcommand{\geant}     {\sc{geant}}
\newcommand{\met}       {\mbox{$\not\!\!E_T$}}
\newcommand{\rar}       {\rightarrow}
\newcommand{\eps}       {\epsilon}
\newcommand{\bs}        {{\it b $\rightarrow$ s$\gamma$}}
\newcommand{\tb}        {{\it t $\rightarrow$ b}}
\newcommand{\coss}      {\mbox{\rm{cos}$\theta^{\star}$}}
\newcommand{\cossp}     {\mbox{\rm{cos}$\theta^{\star}_p$}}
\newcommand{\ptlep}     {$P_T^{lepton}$}
\newcommand{\ljets} {$\ell+$jets}

\title{Measurement of the W Boson Helicity in Top Quark Decay}

%

\author{Amitabha Das}
\affiliation{Department of Physics, University of Arizona, Tucson, AZ, USA}

\begin{abstract}
We present a measurement of the helicity of the W boson produced in top quark decays using \ttbar ~decays in the $\ell+$jets and dilepton final states selected from a sample of 5.4 fb$^{-1}$ of collisions recorded using the D0 detector at the Fermilab Tevatron \ppbar\ collider.  We measure the fractions of longitudinal and right-handed $W$ bosons to be $f_0 = 0.669 \pm 0.078 \hbox{ (stat.)} \pm  0.065 \hbox{ (syst.)}$ and $f_+ = 0.023 \pm 0.041 \hbox{ (stat.)} \pm  0.034  \hbox{ (syst.)}$, respectively.  This result is consistent at the 98\% level with the standard model.  A measurement with $f_0$ fixed to the value from the standard model yields $f_+ = 0.010 \pm 0.022 \hbox{ (stat.)} \pm 0.030 \hbox{ (syst.). }$
\end{abstract}

\maketitle

\thispagestyle{fancy}


\section{Introduction}
The top quark, which is the heaviest known fundamental particle, was discovered in 1995~\cite{cdftopobs,d0topobs} at the Tevatron proton-antiproton collider at Fermilab. The dominant top quark production mode at the Tevatron is $p\bar{p} \rightarrow t\bar{t}X$.  Since the time of discovery, over 100 times more data has been collected, providing a large number of \ttbar\ events with which to study the properties of the top quark. In the standard model (SM), the branching ratio for the top quark to decay to a $W$ boson and a $b$ quark is $> 99.8$\%. The on-shell $W$ boson from the top quark decay has three possible helicity states depending on the relative direction of the spin and momentum of the $W$ boson. We define the fraction of $W$ bosons produced in these states as $f_0$ (longitudinal), $f_-$ (left-handed), and $f_+$ (right-handed). In the SM, the top quark decays via the $V-A$ charged weak current interaction, which strongly suppresses right-handed $W$ bosons. The SM expected values are $f_0$=0.698, $f_-$=0.301, and $f_+=4.1 \times10^{-4}$. The uncertainties on the SM expectations are $\approx (1 - 2)$\% for $f_0$ and $f_-$, and ${\cal O}(10^{-3})$ for $f_+$~\cite{fval}. 

Here we present a measurement of the $W$ boson helicity fractions $f_0$ and $f_+$ and constrain the fraction $f_-$ through the unitarity requirement of $f_- + f_+ + f_0 = 1$. Any significant deviation from the SM expectation would be an indication of new physics, arising from either a deviation  from the expected $V-A$ coupling of the $tWb$ vertex or the presence of non-SM events in the data sample.

The extraction of the $W$ boson helicities is based on the measurement of the angle $\theta^{\star}$ between the directions of the top quark and the down-type fermion (charged lepton or $d$, $s$ quark) decay product of the $W$ boson in the $W$ boson rest frame.  The dependence of the distribution of \coss on the $W$ boson helicity fractions is given by 
\begin{eqnarray}
\omega(c) \propto 2(1-c^2)f_0 + (1-c)^2 f_- + (1+c)^2 f_+
\label{eq:expcost}
\end{eqnarray}
with $c=\coss$. After selection of a ttbar enriched sample the four-momenta of the \ttbar\ decay products in each event are reconstructed as described below, permitting the calculation of \coss. Once the \coss\ distribution is measured, the values of $f_0$ and $f_+$ are extracted with a binned Poisson likelihood fit to the data. The measurement presented here is based on \ppbar ~collisions at a center-of-mass energy $\sqrt s$ = 1.96 TeV corresponding to an integrated luminosity  of 5.4~fb$^{-1}$.

\section{Data and Simulation samples}
This analysis is performed using events collected between April 2002 and June 2009, corresponding to a total integrated luminosity of 5.4 fb$^{-1}$. Analysis of the Run IIa sample, which totals about 1 fb$^{-1}$, was presented in Ref.~\cite{prevd0result}. Here we describe the analysis of the Run IIb data sample and then combine our result with the result from Ref.~\cite{prevd0result} when reporting our measurement from the full data sample.

The Monte Carlo (MC)  simulated samples used for modeling the data are generated with {\sc alpgen}~\cite{ref:alpgen} interfaced to {\sc pythia}~\cite{ref:pythia} for parton shower simulation, passed through a detailed detector simulation based on {\sc geant}~\cite{geant}, and reconstructed using the same algorithms as are used for collider data. For the signal (\ttbar) ~sample, this analysis requires MC samples with  arbitrary non-standard values for the $W$ helicity fractions, while {\sc alpgen}  can only produce linear combinations of $V-A$ and $V+A$ $tWb$ couplings. Hence, for this analysis, we use samples that are either purely $V-A$  or purely   $V+A$, and use a reweighting procedure to form models of arbitrary helicity states. {\sc alpgen} is also used for generating all $V+$jets processes where $V$ represents the vector bosons. {\sc pythia} is used for generating diboson ($WW$, $WZ$, and $ZZ$) backgrounds in the dilepton channels. Background from multijet production is modeled using data.

\section{Analysis}

\subsection{Event Selection}

For this analysis, the selection is done in two steps. In the first step, a loose initial selection using data quality, trigger, object identification, and kinematic criteria is applied to define a sample with the characteristics of \ttbar\ events. Subsequently, a multivariate likelihood discriminant is defined to separate the \ttbar ~signal from the background in the data. We use events in the $\ell+$jets and dilepton \ttbar\ decay channels, which are defined below.

 

In the $\ell+$jets decay $\ttbar ~\rightarrow~W^{+}~W^{-}\bbbar ~\rightarrow~\ell\nu~qq^{'} \bbbar$, events contain one charged lepton (where lepton here refers to an electron or a muon), at least four jets with two of them being $b$ jets, and significant missing transverse energy \met. For the dilepton decay channel,  $\ttbar ~\rightarrow W^{+} W^{-} \bbbar ~\rightarrow \bar{\ell}\nu\ell^\prime\bar{\nu^\prime} \bbbar$, the signature is two leptons of opposite charge, two $b$ jets, and significant \met.

The main sources of background after the initial selection in the $\ell+$jets channel are $W+$jets and multijet production; in the dilepton channels they are $Z$ boson and diboson production as well as multijet and $W$+jets production. The multijet contribution to the $\ell+$jets final states in the initially-selected sample is estimated from data following the method described in Ref.~\cite{matrix}. In the dilepton channels we model the background due to jets being misidentified as isolated leptons using data events where both leptons have the same charge.  This background originates from multijets events with two jets misidentified as leptons and from $W+$jets events with one jets misidentified as a lepton.

\begin{table}[ht]
\caption{\label{tab:optimization} The set of variables chosen for use in $L_t$ for the $e$+jets and $\mu+$ jets channels.  The numbers of background and $t\bar{t}$ events in the initially-selected data, as determined from a fit to the $L_t$ distribution, are also presented.}
\begin{center}
\begin{tabular}{|l|c|c|}
\hline 
\textbf{} & \textbf{$e+$jets} & \textbf{$\mu+$jets} \\
\hline 
Events passing initial selection & 1442 & 1250 \\
\hline
 Variables in best $L_t$ & ${\cal C}$ & ${\cal C}$ \\
 & ${H_T}$ & ${H_T}$ \\
 & ${K_{T\text{min}}^\prime}$ & ${K_{T\text{min}}^\prime}$ \\
 & NN$_{b{\rm avg}}$ & NN$_{b{\rm avg}}$ \\
 & $\mathbf \chi^2_k$ & $h$\\
 & ${m_{jj{\text{min}}}}$ & \\
 & Aplanarity & \\
\hline 
$N$ (\ttbar) & 592.6 $\pm$ 31.8 & 612.7 $\pm$ 31.0 \\
\hline
$N$ ($W+$jets) & 690.2 $\pm$ 21.8 & 579.8 $\pm$ 18.6 \\
\hline 
$N$ (multijet) & 180.3 $\pm$ 9.9 & 6.5 $\pm$ 4.9 \\
\hline 
\end{tabular}
\end{center}
\end{table}

\begin{table}[ht]
\caption{\label{tab:ll_ltfits}The set of variables chosen for use in $L_t$ for the dilepton channels.  The number of background and $t\bar{t}$ events in the initially-selected data, as determined from a fit to the $L_t$ distribution, are also presented.}
\begin{center}
\begin{tabular}{|l|c|c|c|}
\hline 
\textbf{} & \textbf{$e\mu$} & \textbf{$ee$} & \textbf{$\mu\mu$}  \\
\hline 
Events passing initial selection & 323 & 3275 & 5740 \\
\hline
 Variables in best $L_t$ & ${\cal A}$,${\cal S}$,$h$,$m_{jj\text{min}}$ & ${\cal A}$,${\cal S}$,$m_{jj\text{min}}$ &  ${\cal A}$,${\cal S}$,$m_{jj\text{min}}$,$K_{T\text{min}}^\prime$ \\
 & $K_{T\text{min}}^\prime$,\met,NN$_{b1}$,$m_{\ell\ell}$ & \met,NN$_{b1}$,$m_{\ell\ell}$ & $\chi^2_Z$,NN$_{b1}$ \\
\hline 
$N$ (\ttbar) & 178.7 $\pm$ 15.6 & 74.9 $\pm$ 10.7 & 86.0 $\pm$ 13.8 \\
\hline
$N$ (background) & 144.3 $\pm$ 14.5 & 3200 $\pm$ 75 & 5654 $\pm$ 76 \\
\hline 
\end{tabular}
\end{center}
\end{table}

To separate the \ttbar\ signal from these sources of background, we define a multivariate likelihood and retain only events above a certain threshold in the value of that likelihood.  The set of variables used in the likelihood and the threshold value are optimized separately for each \ttbar\ decay channel.  The first step in the optimization procedure is to identify a set of candidate variables that may be used in the likelihood. We start with twelve variabels namely, Aplanarity, Sphericity, $H_T$, Centrality, ${K_{T\text{\bf min}}^\prime}$, ${m_{jj{\text{min}}}}$, h, ${ \chi^2_k}$, ${\Delta\phi(\hbox{lepton}, \met)}$, b jet content of the event, \met\ or ${\chi^2_Z}$ and Di-lepton mass ${m_{\ell\ell}}$. We consider all combinations of the above variables to select the optimal set to use for each \ttbar\ decay channel.  For a given combination of variables, the likelihood ratio $L_t$ is defined as

\begin{eqnarray}
L_t = \frac{\exp\left\{\sum_{i=1}^{N_{\rm var}} [\ln(\frac{S}{B})_i^{\text{fit}}]\right\}}
{\exp\left\{\sum_{i=1}^{N_{\rm var}} [\ln(\frac{S}{B})_i^{\text{fit}}]\right\}+ 1}
\label{eq:claslhood}
\end{eqnarray}
where $N_{\rm var}$ is the number of input variables used in the likelihood, and $(\frac{S}{B})_i^{\text{fit}}$ is the ratio of the parameterized signal and background probability density functions. We consider all possible subsets of the above variables to be used in $L_t$ and scan across all potential selection criteria on $L_t$.  For each $L_t$ definition and prospective selection criterion, we compute the following  figure of merit (FOM):
\begin{eqnarray}
  {\rm FOM} = \frac{N_S}{\sqrt{N_S + N_B + \sigma^2_{B}}}
\label{eq:FOM}
\end{eqnarray}
where $N_S$ and $N_B$ are the numbers of signal and background events expected to satisfy the $L_t$ selection. The term $\sigma_{B}$ reflects the uncertainty in the background selection efficiency arising from any mis-modeling of the input variables in our MC.

\begin{figure}
\includegraphics[scale=0.40]{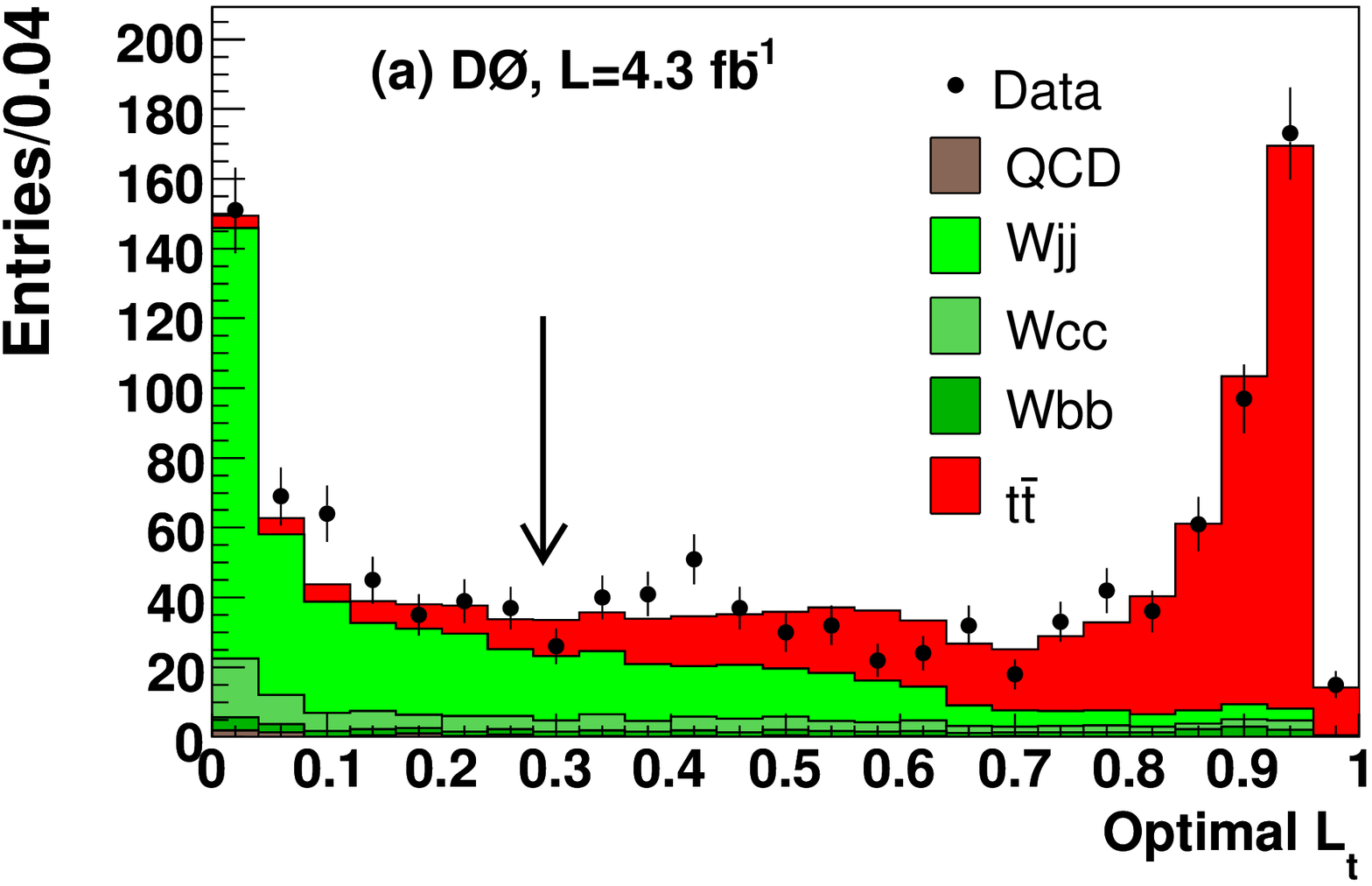}
\includegraphics[scale=0.40]{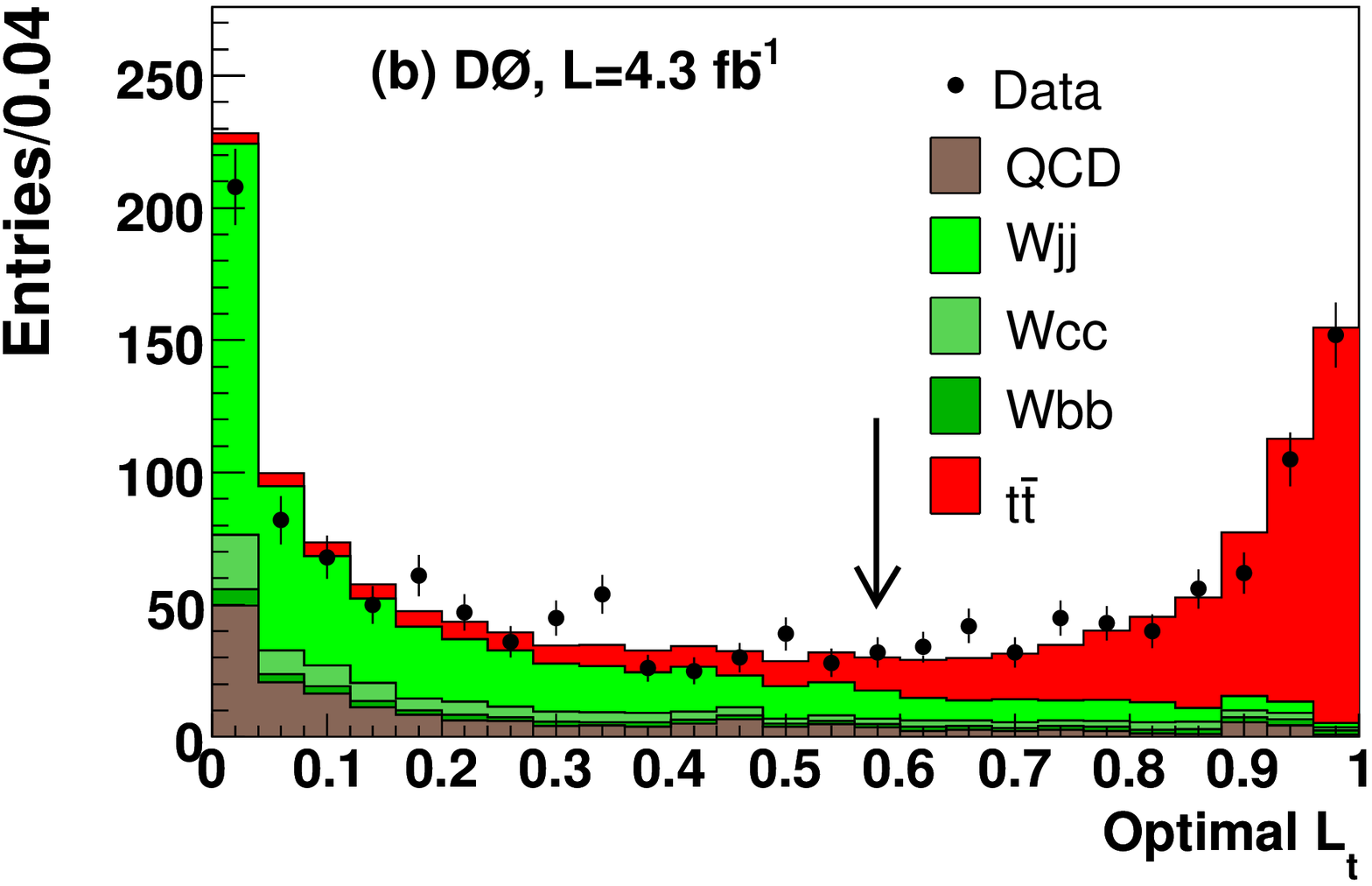}
\caption{\label{fig:BestLt} (Color online) Best $L_t$ variable for the (a) $\mu+$jets and (b) $e+$jets channels. The MC is normalized using the signal and background fractions returned by the Poisson maximum likelihood fit. The arrows mark the required $L_t$ values for events in each channel.}
\end{figure}

\begin{figure}
\includegraphics[scale=0.40]{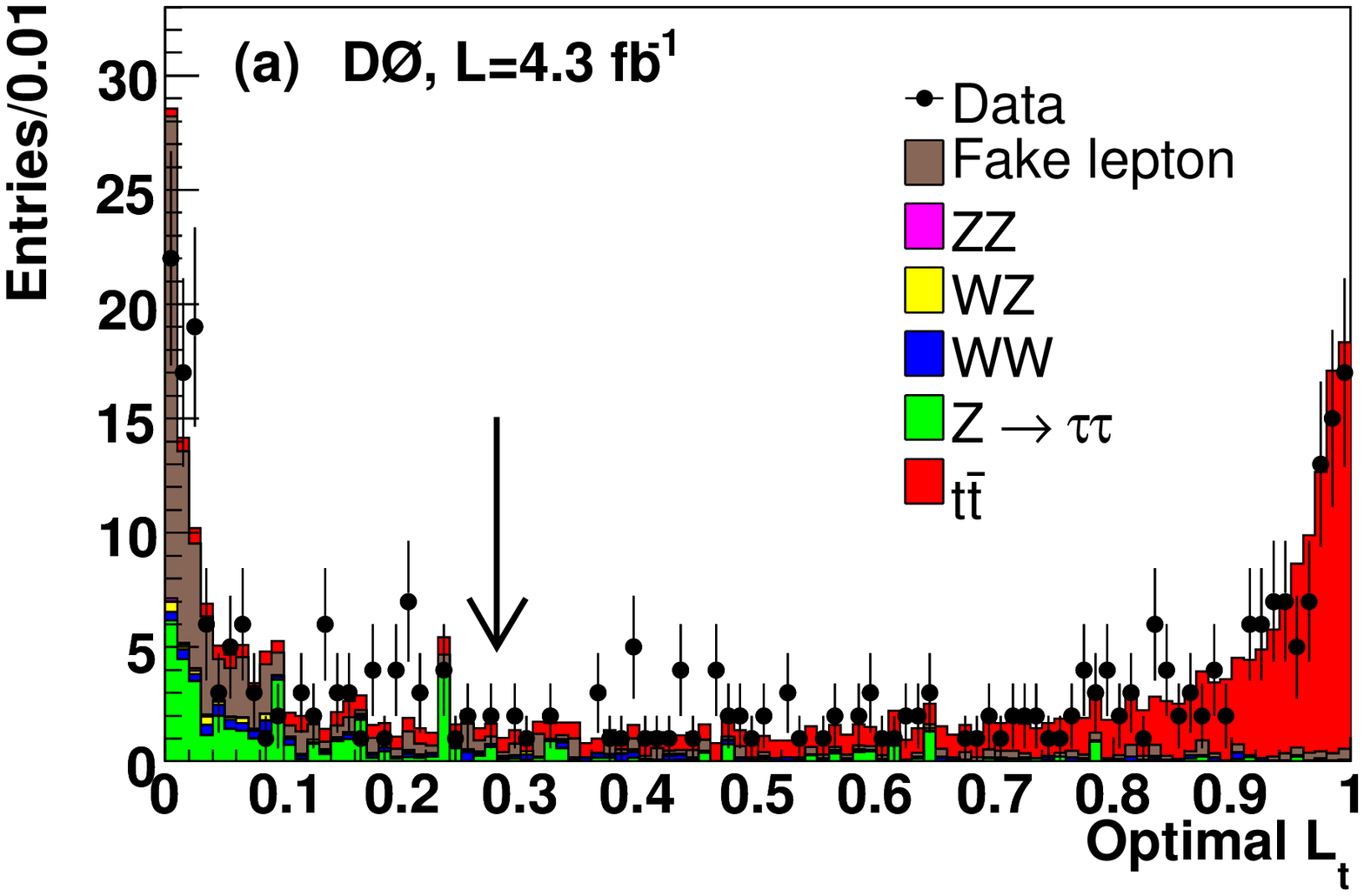}
\includegraphics[scale=0.40]{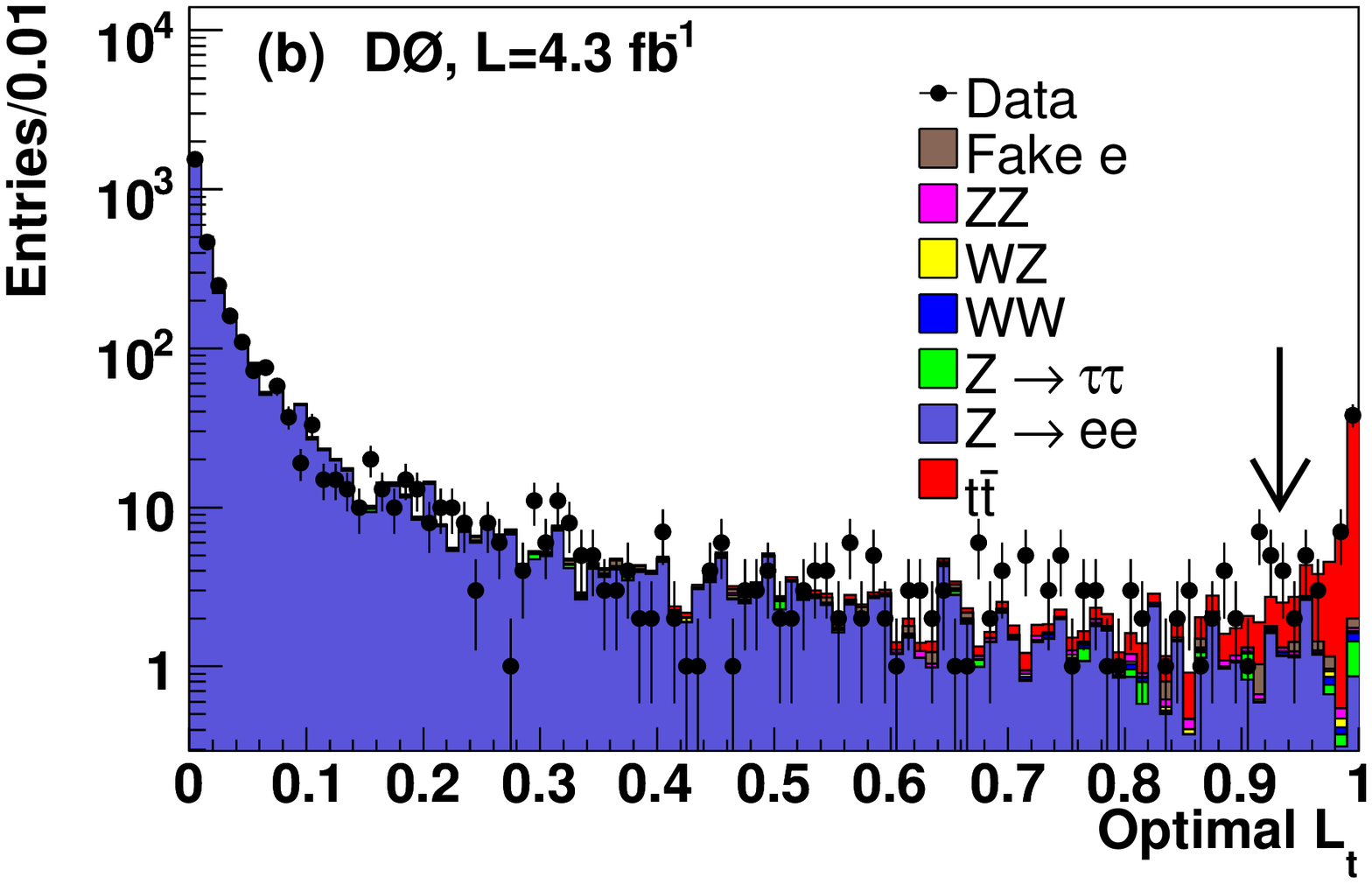}
\includegraphics[scale=0.40]{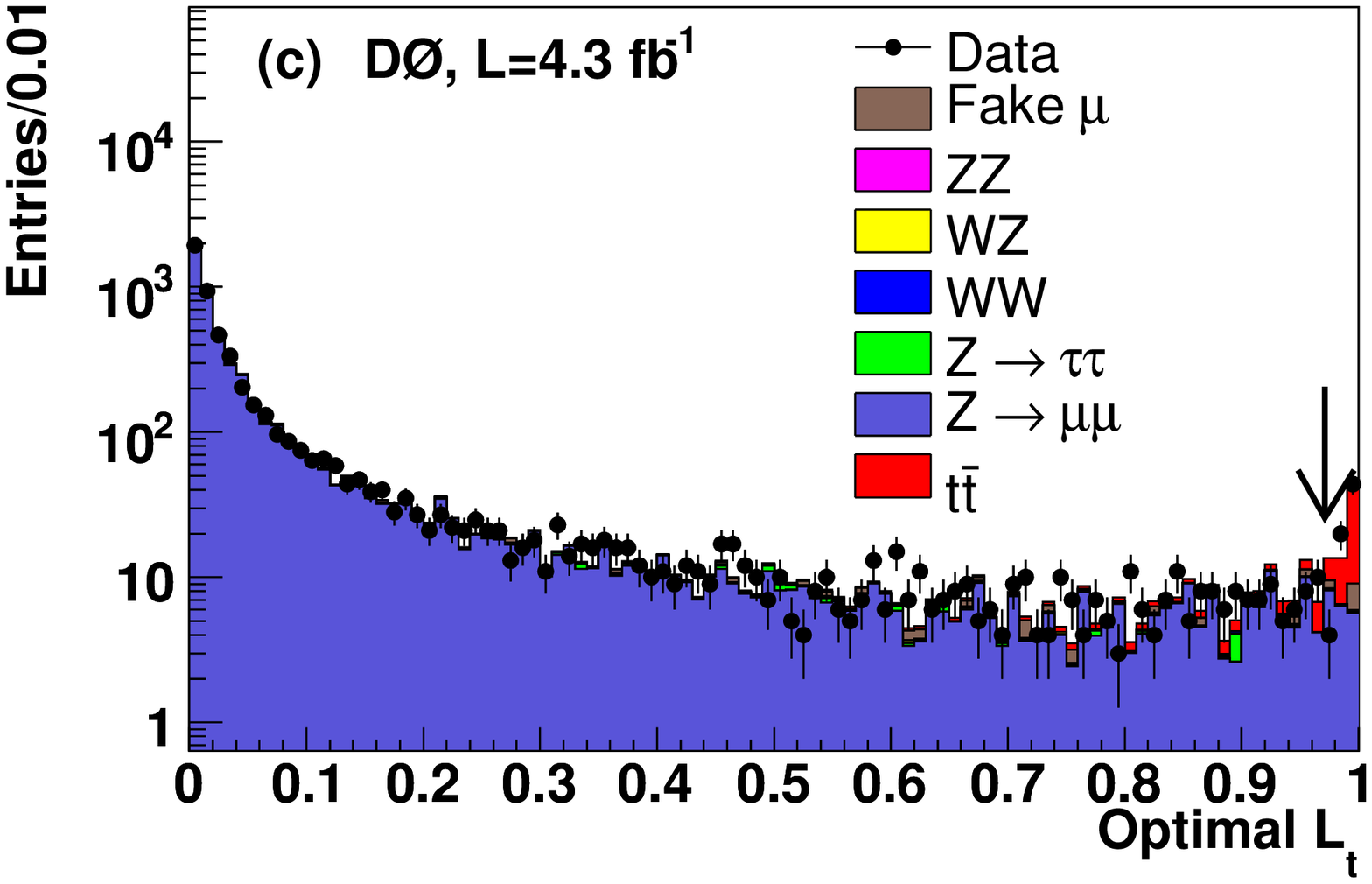}
\caption{\label{fig:BestLtll} (Color online) Best $L_t$ variable for the (a) $e\mu$, (b) $ee$  and (c) $\mu\mu$ decay channels. The MC is normalized using the signal and background fractions returned by the Poisson maximum likelihood fit to the $L_t$ distribution.  The arrows mark the required $L_t$ values for events in each channel.}
\end{figure}

The signal and background yields in the initially-selected sample for the $\mathbf{\ell+}$jets channels are listed in Table~\ref{tab:optimization}, and for the dilepton channels in Table~\ref{tab:ll_ltfits}. Tables~\ref{tab:data_selection} and~\ref{tab:llfinal} show the optimal $L_t$ cut value for each channel and the final number of events in data and the expected numbers of signal and background events after applying the $L_t$ requirement. Figures~\ref{fig:BestLt} and~\ref{fig:BestLtll} show the distribution of the best likelihood discriminant for each channel, where the signal and background contributions are normalized according to the values returned by the fit.

\begin{table}[ht]
\caption{\label{tab:data_selection} Expected background and $t\bar{t}$ yields, and the number of events observed, after the selection on $L_t$ in the $\ell+$jets decay channels.}
\begin{center}
\begin{tabular}{|l|c|c|}
\hline 
\textbf{} & \textbf{$e+$jets} & \textbf{$\mu+$jets} \\
\hline 
Best $L_t$ cut & 0.58 & 0.29 \\
\hline
Expected \ttbar                   & 484.4 $\pm$ 41.4 & 567.2 $\pm$ 47.3\\
Expected $W+$jets                 & 111.7 $\pm$ 12.6 & 227.7 $\pm$ 19.2\\
Expected multijet                      &  58.1 $\pm$ 3.9 & 4.0 $\pm$ 3.1\\ \hline
Expected total                 &  656.2 $\pm$ 43.4 & 798.9 $\pm$ 51.2\\ \hline
Observed                           &  628 & 803 \\
\hline 
\end{tabular}
\end{center}
\end{table}

\begin{table}[ht]
\caption{\label{tab:llfinal}  Expected background and $t\bar{t}$ yields, and the number of events observed, after the selection on $L_t$ in the dilepton decay channels.}
\begin{center}
\begin{tabular}{|l|c|c|c|}
\hline 
\textbf{} & \textbf{$e\mu$} & \textbf{$ee$} & \textbf{$\mu\mu$} \\
\hline 
Best $L_t$ cut & $>$ 0.28 & $>$ 0.934 & $>$ 0.972 \\
\hline
Expected $t\bar{t}$ & 186.6 $\pm$ 0.4 & 44.5 $\pm$ 0.3  &  43.6 $\pm$ 0.3  \\ 
Expected $Z/\gamma^* \rightarrow \ell^+\ell^-$  & N/A & 7.4 $\pm$ 1.0 &  19.1 $\pm$ 1.3\\
Expected $Z/\gamma^* \rightarrow \tau\tau$      & 11.2  $\pm$ 3.7  & 0.8 $\pm$ 0.3 & 0.35 $\pm$ 0.05 \\
Expected $WW$                          & 5.6 $\pm$ 1.4   & 0.3 $\pm$  0.1& 0.13 $\pm$ 0.05\\
Expected $WZ$                          &  1.5 $\pm$ 0.5 & 0.28 $\pm$ 0.04 & 0.16 $\pm$ 0.01\\
Expected $ZZ$                          &  1.0 $\pm$ 0.5 & 0.34 $\pm$ 0.04 & 0.57 $\pm$ 0.04\\
Expected misidentified jets                   & 15.9 $\pm$ 3.1       & 0.54 $\pm$ 0.48 & 3.7 $\pm$ 2.5\\ \hline
Expected total & 221.7 $\pm$ 5.1 & 54.2 $\pm$ 1.2 & 67.7 $\pm$ 3.9 \\ \hline
Observed &  193   & 58  &  68 \\ \hline
\end{tabular}
\end{center}
\end{table}

\subsection{Templates}

After the final event selection if performed, \coss\ is calculated for each event by using the reconstructed top quark and $W$ boson four-momenta. In the $\ell+$jets decay channel, the four-momenta are reconstructed using a kinemetic fit with the constraints: (i) two jets should give the invariant mass of the $W$ boson (80.4 GeV/$c^2$), (ii) one lepton and the \met\ should give the invariant mass of the $W$ boson, and (iii) the mass of the reconstructed top and anti-top quark should be 172.5 GeV/$c^2$. In the $\ell+$jets decay channel, the hadronic $W$ boson decay from the top quark in the event also contains information about the helicity of that $W$ boson. Since we do not distinguish between jets formed from up-type and down-type quarks, we choose one of the $W$ boson daughter jets at random as the basis for the calculation. With this choice, left-handed and right-handed $W$ bosons have identical $|\coss|$ distributions, but we can distinguish either of those states from longitudinal $W$ bosons, thereby improving the precision of our measurement.  

\subsection{Model-independent $W$ Helicity Fit}

\begin{figure}
\includegraphics[scale=0.45]{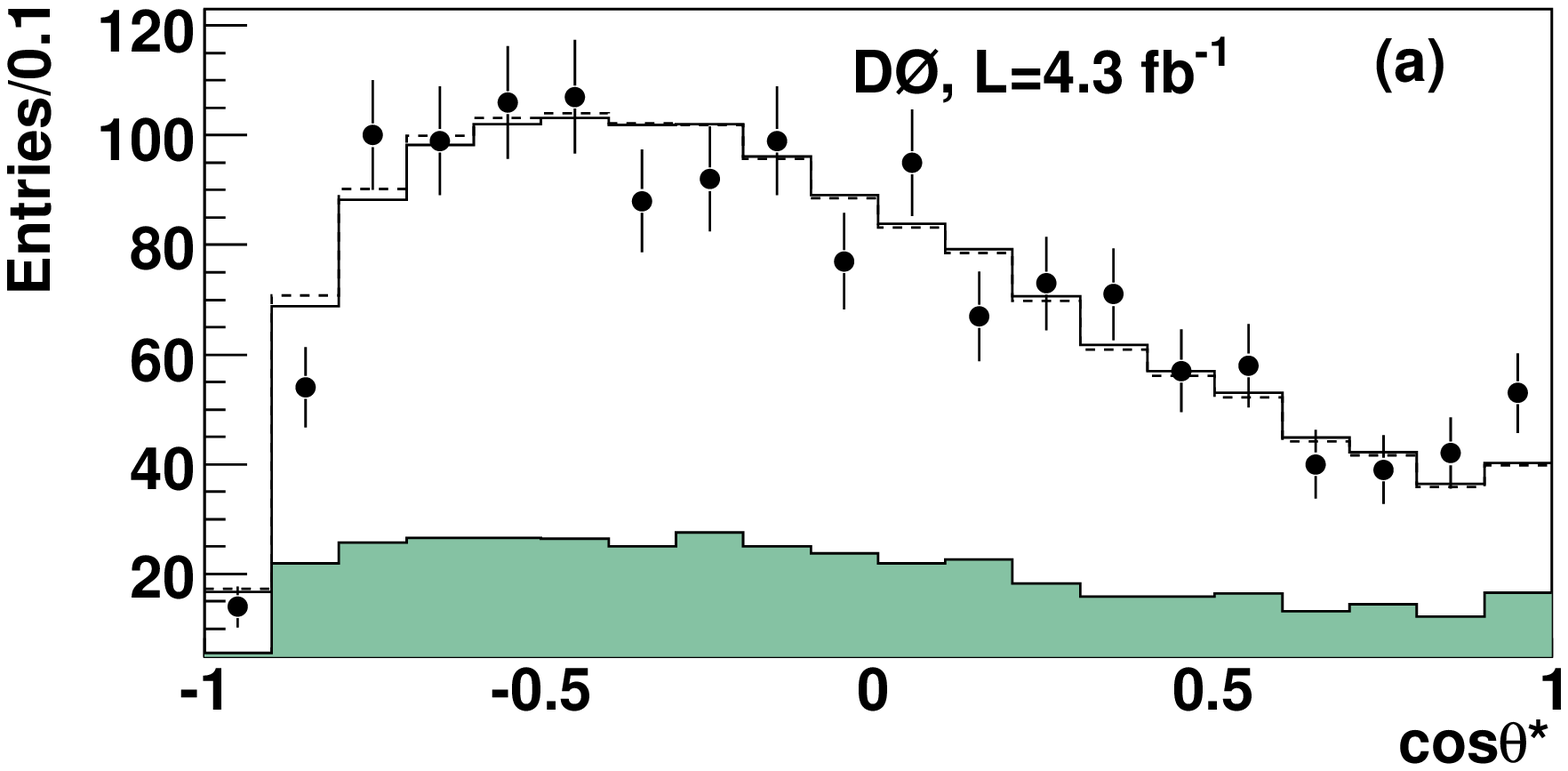} \\
\includegraphics[scale=0.45]{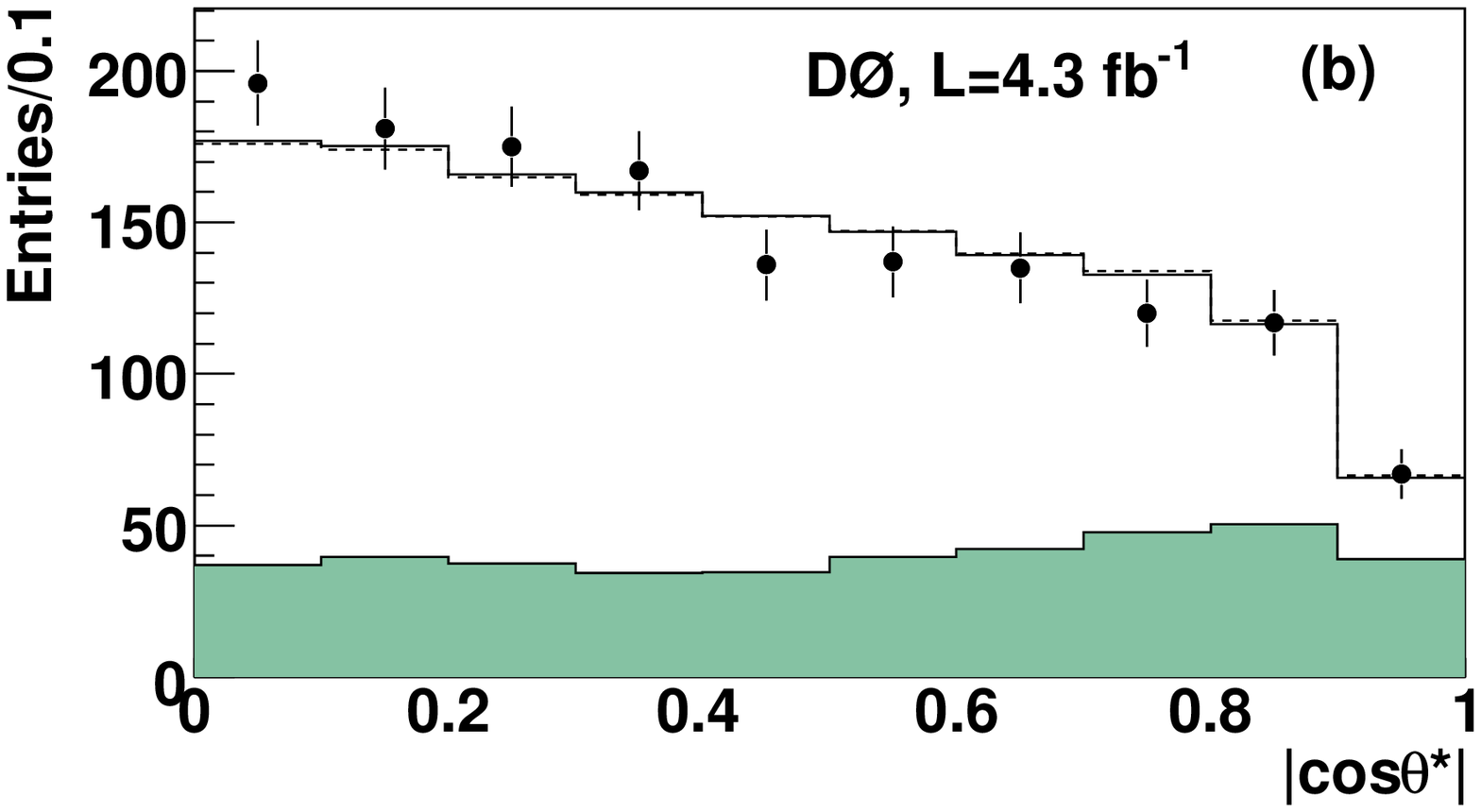}\\
\includegraphics[scale=0.45]{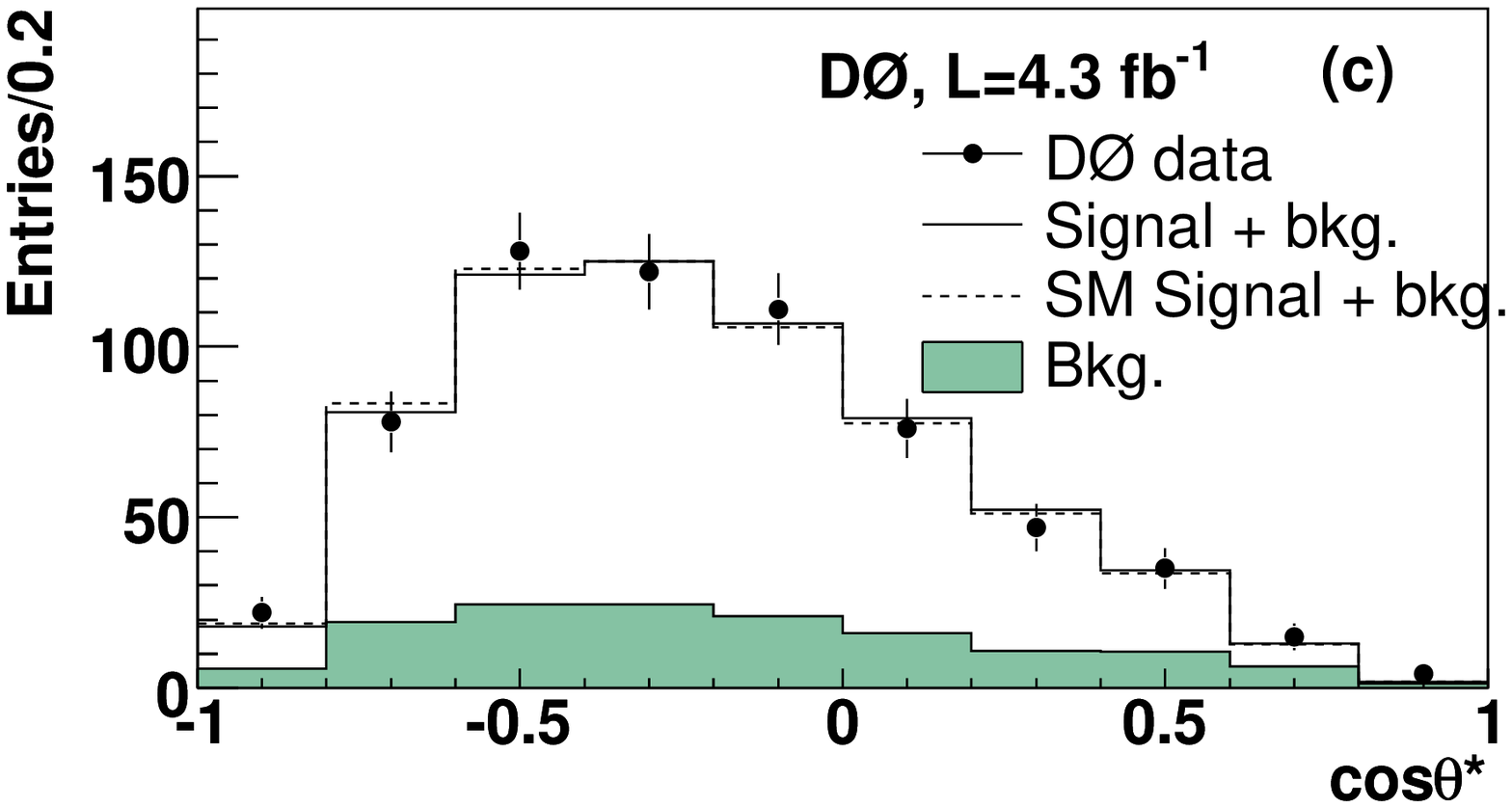}
\caption{\label{fig:data2dmodel} 
(Color online) Comparison of the \coss\ distribution in Run IIb data and the global best-fit model (solid line) and the SM (dashed line) for (a) leptonic $W$ boson decays in \ljets\ events, (b) hadronic $W$ boson decays in \ljets\ events, and (c) dilepton events.}
\end{figure}

The $W$ boson helicity fractions are extracted by computing a binned Poisson likelihood $L(f_0,f_+)$ with the distribution of \coss\ in the data to be consistent with the sum of signal and background templates. The likelihood is a function of the $W$ boson helicity fractions $f_0$ and $f_+$. While performing the fit, both $f_0$ and $f_+$ are allowed to float freely, and the measured $W$ helicity fractions correspond to those leading to the highest likelihood value. The comparison between the best-fit model and the data is shown in Fig.~\ref{fig:data2dmodel}.

\subsection{Systematics}

Systematic uncertainties are evaluated using simulated event ensembles in which both changes in the background yield and changes in the shape of the \coss\ templates in signal and background are considered. The simulated samples from which the events are drawn can be either our nominal samples or samples in which the systematic effect under study has been shifted away from the nominal value. In general, the systematic uncertainties assigned to $f_0$ and $f_+$ are determined by taking an average of the absolute values of the differences in the average fit output values between the nominal and shifted $V-A$ and $V+A$ samples.

\begin{table}[hhh]
\caption{\label{tab:2dsyst} Summary of the systematic uncertainties on $f_+$ and $f_{0}$.}
\begin{tabular}{|l|c|c|}
\hline
Source & Uncertainty ($f_+$) & Uncertainty ($f_0$)  \\ \hline
Jet energy scale       & 0.007  & 0.009 \\
Jet energy resolution  & 0.004 & 0.009 \\
Jet ID        & 0.004 & 0.004 \\
Top quark mass          & 0.011 & 0.009 \\
Template statistics   & 0.012 & 0.023 \\
\ttbar ~model          & 0.022 & 0.033 \\
Background model       & 0.006   & 0.017  \\
Heavy flavor fraction   & 0.011 & 0.026 \\
$b$ fragmentation      & 0.000 &  0.001 \\
PDF                 & 0.000 & 0.000 \\
Analysis consistency   & 0.004 &  0.006\\
Muon ID            &  0.003 & 0.021 \\
Muon trigger     &  0.004 & 0.020 \\  \hline
Total                  & 0.032 &   0.060 \\ \hline
\end{tabular}
\end{table}

\subsection{Results}

Applying the model independent fit to the Run IIb data, we find
\begin{eqnarray}
f_0 &=& 0.739 \pm 0.091 \hbox{ (stat.)} \pm  0.060  \hbox{ (syst.)} \\
f_+ &=& -0.002 \pm 0.045 \hbox{ (stat.)} \pm  0.032 \hbox{ (syst.)}.
\end{eqnarray}

The 68\% and 95\% C.L. contours in the $(f_+,f_0)$ plane are shown in Fig.~\ref{fig:data2dfit} (a). Finally, we perform fits in which one of the two helicity fractions is fixed to its SM value. Constraining $f_0$, we find
\begin{eqnarray}
f_+ = 0.014 \pm 0.025 \pm \hbox{ (stat.)} \pm 0.028  \hbox{(syst.)} ,
\end{eqnarray}

We also constrain $f_+$ and measure $f_0$, finding
\begin{eqnarray}
f_0 = 0.735 \pm 0.051 \hbox{ (stat.)} \pm 0.051   \hbox{(syst.)}.
\end{eqnarray}

\subsection{Combination with Our Previous Measurement}

To combine this result with the previous measurement from Ref.~\cite{prevd0result}, we repeat the maximum likelihood fit with the earlier and current data samples and their respective MC models, treating them as separate channels in the fit.  This is equivalent to multiplying the two-dimensional likelihood distributions in $f_0$ and $f_+$ corresponding to the two data sets.  We determine the systematic uncertainty on the combined result by treating most uncertainties as correlated (the exception is template statistics) and propagating the uncertainties to the combined result.  The results are presented in Table~\ref{tab:p17p20combsyst}.

\begin{table}[hhh]
\caption{\label{tab:p17p20combsyst} Summary of the combined systematic uncertainties on $f_+$ and $f_{0}$ for Run IIa and Run IIb.}
\begin{tabular}{|l|c|c|}
\hline
Source & Uncertainty ($f_+$) & Uncertainty ($f_0$)  \\ \hline
Jet energy scale       & 0.009  & 0.010 \\
Jet energy resolution  & 0.004 & 0.008 \\
Jet ID        & 0.005 & 0.007 \\
Top mass          & 0.012 & 0.009\\
Template statistics   & 0.011 & 0.021 \\
\ttbar ~model          & 0.024 & 0.039 \\
Background model       & 0.008   & 0.023  \\
Heavy flavor fraction   & 0.010 & 0.022 \\
$b$ fragmentation  & 0.002 &  0.004 \\
PDF                 & 0.000 & 0.001 \\
Analysis consistency   & 0.004 & 0.006 \\
Muon ID         &   0.002 & 0.017 \\
Muon trigger   &  0.003 & 0.024 \\ \hline
Total                  & 0.034  &   0.065 \\ \hline
\end{tabular}
\end{table}

\begin{figure*}
\includegraphics[scale=0.40]{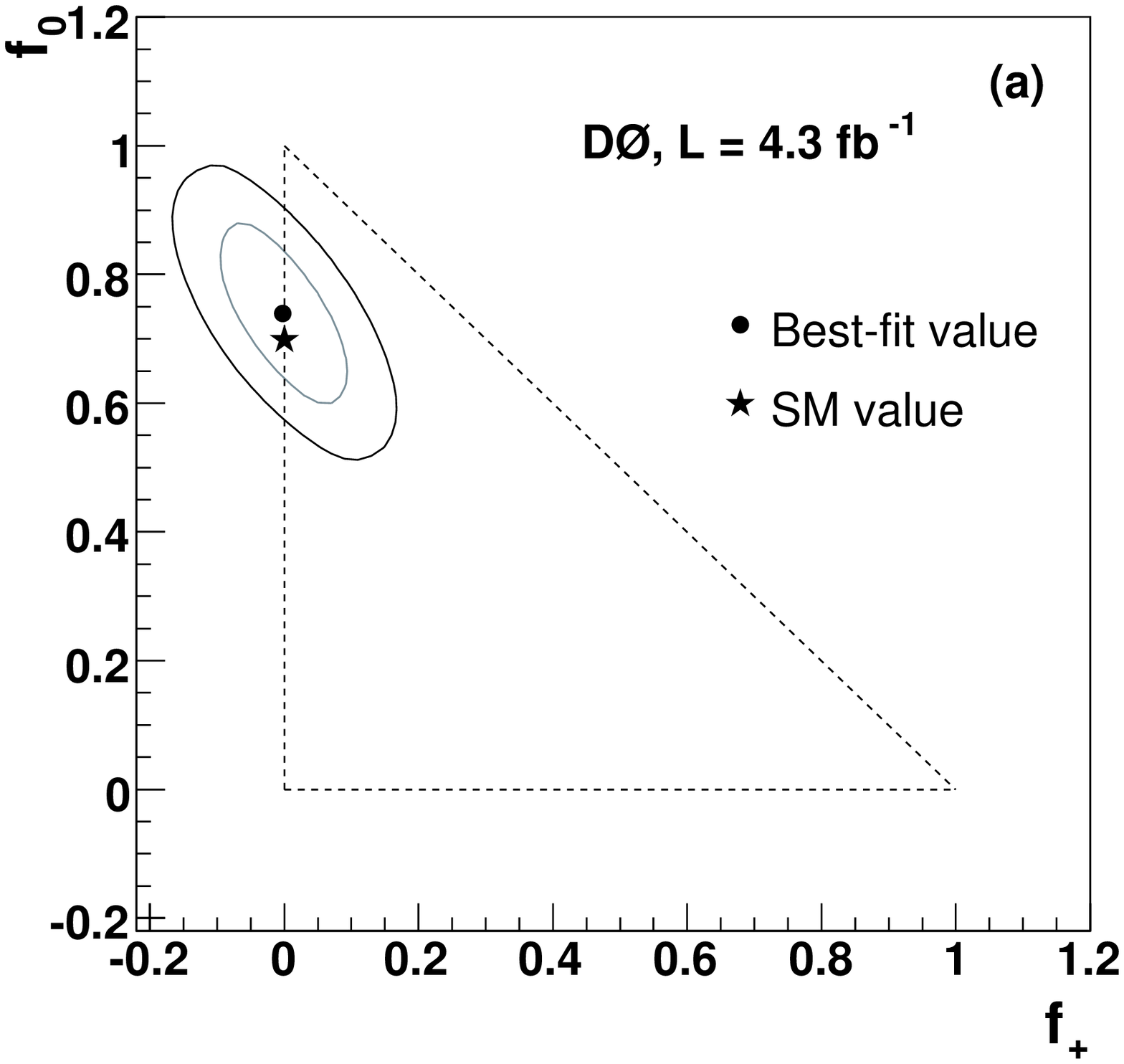}
\includegraphics[scale=0.40]{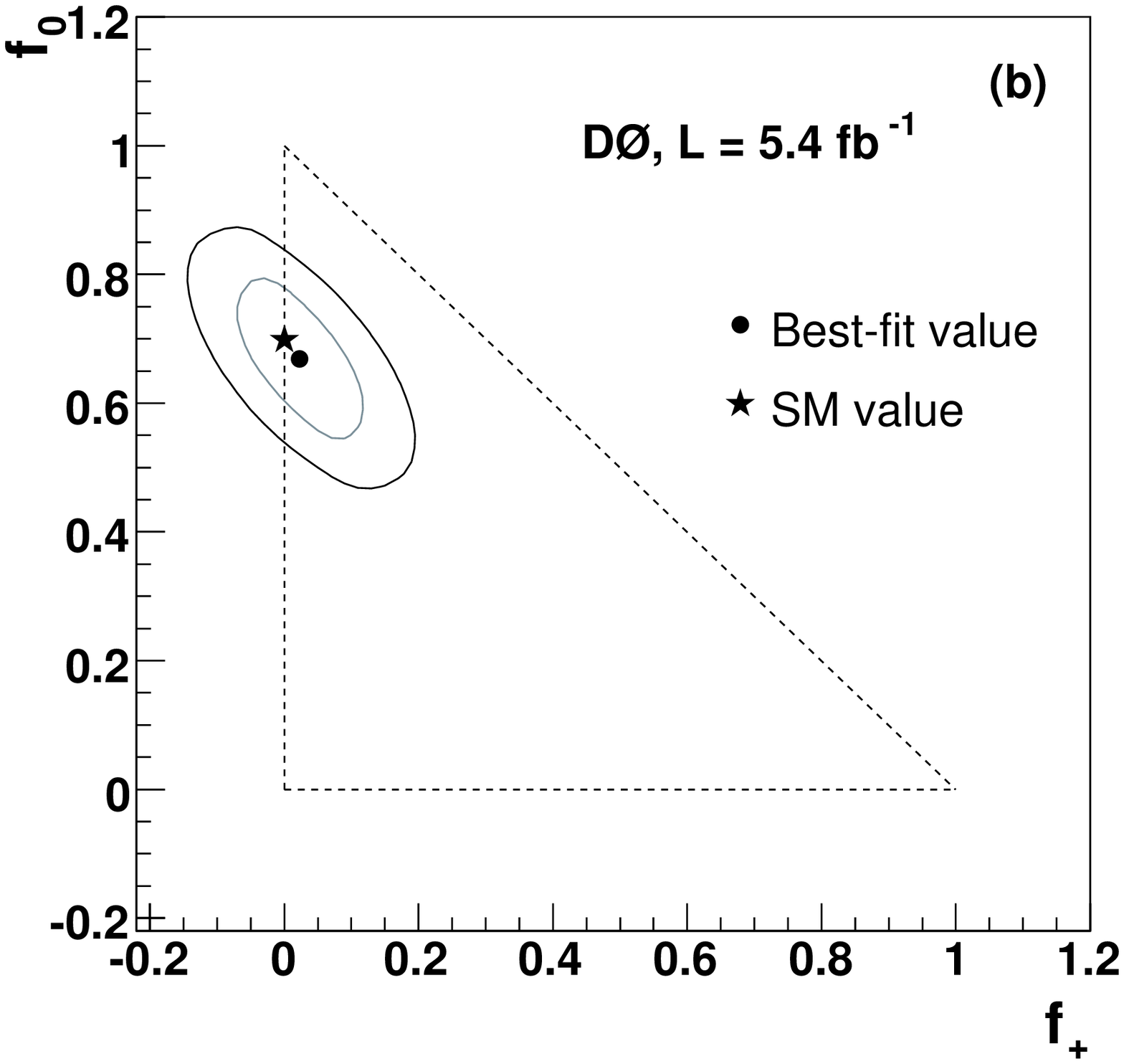}
\caption{\label{fig:data2dfit}  Result of the model-independent $W$ boson helicity fit for (a) the Run IIb data sample and (b) the combined Run IIa and Run IIb data sample.  In both plots, the ellipses indicate the 68\% and  95\% C.L.  contours, the dot shows the best-fit value, the triangle corresponds to the physically allowed region where $f_0 + f_+ \le 1$, and the star marks the expectation from the SM.}
\end{figure*}

The combined result for the entire 5.4 fb$^{-1}$ sample is
\begin{eqnarray}
f_0 &=& 0.669 \pm 0.078 \hbox{ (stat.)} \pm  0.065 \hbox{ (syst.)} \\
f_+ &=& 0.023 \pm 0.041 \hbox{ (stat.)} \pm  0.034 \hbox{ (syst.)}.
\end{eqnarray}
The 68\% and 95\% C.L. contours in the $(f_+,f_0)$ plane are shown in Fig.~\ref{fig:data2dfit} (b). The probability of observing a greater deviation from the SM due to random chance is 83\% when only statistical uncertainties are considered and 98\% when systematic uncertainties are included.

Constraining $f_0$ to be 0.7,  we find
\begin{eqnarray}
f_+ = 0.010 \pm 0.022 \hbox{ (stat.)} \pm 0.030  \hbox{ (syst.)}
\end{eqnarray}

Constraining $f_+$ to 0 gives
\begin{eqnarray}
f_0 = 0.708 \pm 0.044 \hbox{ (stat.)} \pm 0.048 \hbox{ (syst.)}
\end{eqnarray}
 
\section{Conclusion}
We have measured the helicity of $W$ bosons arising from top quark decay in \ttbar\ events using both the $\ell+$jets and dilepton decay channels and find
\begin{align}
f_0 = 0.669 & \pm  0.102  [ \pm  0.078 \hbox{ (stat.)} \pm  0.065 \hbox{ (syst.)}], \nonumber \\
\\ 
f_+ = 0.023 & \pm   0.053  [ \pm  0.041 \hbox{ (stat.)} \pm  0.034  \hbox{ (syst.)}]. \nonumber
\end{align}
in a model-independent fit.  The consistency of this measurement with the SM values $f_0 = 0.698$, $f_+=3.6\times10^{-4}$ is 98\%.  Therefore, we report no evidence for new physics at the $tWb$ decay vertex.

\begin{acknowledgments}
We thank the staffs at Fermilab and collaborating institutions, 
and acknowledge support from the
DOE and NSF (USA);
CEA and CNRS/IN2P3 (France);
FASI, Rosatom and RFBR (Russia);
CNPq, FAPERJ, FAPESP and FUNDUNESP (Brazil);
DAE and DST (India);
Colciencias (Colombia);
CONACyT (Mexico);
KRF and KOSEF (Korea);
CONICET and UBACyT (Argentina);
FOM (The Netherlands);
STFC and the Royal Society (United Kingdom);
MSMT and GACR (Czech Republic);
CRC Program and NSERC (Canada);
BMBF and DFG (Germany);
SFI (Ireland);
The Swedish Research Council (Sweden);
and
CAS and CNSF (China).
\end{acknowledgments}

\bigskip 

\end{document}